\pdfoutput=1
\documentclass[nohyper,12pt,a4paper]{JHEP3}
\usepackage{epsfig}
\usepackage{psfrag}
\usepackage{graphicx}
\usepackage{amsfonts,amsmath,amssymb}

\title{Holographic metals at finite temperature}

\author{ V. Giangreco M. Puletti,$^{1)}$ S. Nowling,$^{1)}$ L. Thorlacius,$^{1),2)}$ and 
T. Zingg$^{1),2)}$\\ \ \\
1) NORDITA, Roslagstullsbacken 23 \\
SE-106 91 Stockholm, Sweden \\ 
\ \\
2) University of Iceland, Science Institute \\
Dunhaga 3, IS-107 Reykjavik, Iceland\\
\ \\
E-mail:  \email{valentina}, \email{nowling},
\email{larus}, \email{zingg} \email{ @nordita.org}}

\abstract{A holographic dual description of a 2+1 dimensional system of strongly 
interacting fermions at low temperature and finite charge density is given in terms 
of an electron cloud suspended over the horizon of a charged black hole in 
asymptotically AdS spacetime. The electron star of Hartnoll and Tavanfar is 
recovered in the limit of zero temperature, while at higher temperatures the 
fraction of charge carried by the electron cloud is reduced and at a critical 
temperature there is a third order phase transition to a configuration with 
only a charged black hole. The geometric structure implies that finite 
temperature transport coefficients, including the AC electrical conductivity, 
only receive contributions from bulk fermions within a finite band in the radial
direction.}

\preprint{NORDITA-2010-102, RH-27-2010}

\begin{document}


\section{Introduction}

There has been considerable recent interest in developing holographic models 
of strongly coupled physics in low dimensions with a view towards condensed
matter systems 
(see \cite{Herzog:2009xv,Hartnoll:2009sz,McGreevy:2009xe,Sachdev:2010uj}
for reviews). This can be motivated both from the point of view of extending
the gauge theory/gravity correspondence to include a variety of interesting 
field theoretic systems without supersymmetry, and also from the point of view 
of gaining a new theoretical handle on materials containing strongly correlated 
electrons. 

In a recent paper Hartnoll and Tavanfar \cite{Hartnoll:2010gu} considered a 
simple holographic model for strongly interacting fermions in 2+1 dimensions
at zero temperature and finite charge density. In their approach, which builds
on earlier work in 
\cite{deBoer:2009wk,Hartnoll:2009ns}, the bulk  Maxwell field, 
which is dual to the field theory current, is sourced by an
ideal fluid consisting of charged free fermions. The combined Einstein,
Maxwell, and fluid field equations in the bulk spacetime have planar
solutions, referred to as electron stars in \cite{Hartnoll:2010gu}, where
the charge and energy densities of the bulk fermion fluid have a non-trivial 
radial profile. The geometry is asymptotically AdS but deep in the electron
star interior the metric exhibits Lifshitz scaling with a non-universal 
dynamical critical exponent that depends on the couplings of the model.
Similar constructions were considered in \cite{Arsiwalla:2010bt, Parente:2010fs} for neutral 
and charged free fermion fluids supported by a degeneracy pressure. 

In the present paper we extend the gravity dual description 
of \cite{Hartnoll:2010gu} to include finite 
temperature configurations in the boundary field theory. We construct static
solutions of the bulk field equations where an `electron cloud' is suspended
above the horizon of a charged black hole, or more precisely a black brane
with a planar horizon. The electron cloud has both
an outer and an inner edge. The outer edge is also found in the electron 
stars of \cite{Hartnoll:2010gu} but the inner edge is a new feature, found
only at finite temperature. At the inner edge the gravitational pull of the
black brane on the electron fluid is balanced by electrostatic repulsion.  

In the fluid description, observables in the boundary 
theory, such as the electric conductivity, only receive contributions from 
bulk fermions located within a band of finite width in the radial direction. 
The sharpness of the edges is presumably an artifact of the classical
perfect fluid description and we expect both quantum corrections and 
fluid interactions to give rise to tails in the bulk fermion profile that fall 
off towards the boundary and the 
black brane horizon, respectively.\footnote{Such tails are, for instance,
found in an alternative approach to including bulk fermions based on a 
single fermion wave equation~\cite{Cubrovic:2010bf}.}

At the level of classical geometry, a finite temperature in the boundary 
theory is introduced by including a black hole in the bulk spacetime, with
a non-vanishing Hawking temperature. Quantum effects in the bulk 
include thermal Hawking radiation, which comes to equilibrium with the 
bulk fermion fluid, but at weak gravitational coupling this thermalization 
in the bulk is suppressed and will not be considered here. This simplifies
our analysis considerably as it allows us to use a zero temperature 
equation of state for the free fermions in the bulk to capture finite 
temperature effects in the boundary field theory.

At low temperatures in the boundary theory, the electric charge in the bulk 
geometry is partly carried by the electron cloud and partly by the charged 
black brane inside it. As the temperature is raised, the two edges of the electron 
cloud move towards each other and an ever larger fraction of the total electric 
charge in the bulk geometry resides inside the black brane. The two edges
of the electron fluid meet at a finite critical temperature, above which there 
is only a black brane solution with no electron cloud present. We find that 
the system undergoes a third order phase transition at the critical point.\footnote{In an earlier preprint of this paper it was incorrectly stated that the phase transition was second order.  The correct behavior was identified in \cite{Hartnoll:2010ik}.} 

As the temperature is lowered, on the other hand, the inner edge of the 
electron cloud approaches the black brane horizon, which in turn recedes 
towards vanishing area. In the zero temperature limit, one recovers the electron star geometry where there is no longer any black 
hole and all the charge is carried by the electron fluid. We confirm this by showing that the horizon recedes from the boundary, the geometry at low temperature encodes the dynamical exponent $z$, and by comparing free energy densities. We find that at low 
temperatures the free energy density of an electron cloud geometry 
smoothly goes over to that of an electron star. Furthermore, we compare the free energy densities between an electron cloud solution and
an AdS-RN black brane solution at finite temperature.  We find that, 
whenever a solution with an electron cloud exists, it is favored over an 
AdS-RN black brane. 

The electrical conductivity at finite temperature can be obtained for this
system using, by now, standard holographic techniques (see for instance~\cite{Herzog:2009xv, Hartnoll:2009sz}). We find that the finite temperature conductivity smoothly interpolates between the AdS-RN and electron star results~\cite{Hartnoll:2010gu}. 

The same finite temperature solutions were found independently by Hartnoll and Petrov in \cite{Hartnoll:2010ik}.  Initially there was a discrepancy between their work and ours in the analysis of the phase transition, but after correcting an error in our expression for the free energy density, we now also find a third order phase transition.

\section{Field equations and electron cloud solutions
\label{setup}}

The Einstein-Maxwell equations with a negative cosmological constant 
and a charged perfect fluid are 
\begin{equation}\label{fieldeqs}
R_{\mu\nu}-\frac{1}{2}g_{\mu\nu}R-\frac{3}{L^2}g_{\mu\nu} =
\kappa^2(T^\textrm{Maxwell}_{\mu\nu}+T^\textrm{fluid}_{\mu\nu}) \,, 
\qquad
\nabla^\nu F_{\mu\nu}=e^2 \, J^\textrm{fluid}_\mu \, .
\end{equation}
We adopt units where the characteristic AdS length scale is $L=1$.  
The source terms are given by
\begin{eqnarray}
T^\textrm{Maxwell}_{\mu\nu}&=& 
\frac{1}{e^2}(F_{\mu\lambda}F_\nu^{\ \lambda}
-\frac{1}{4}g_{\mu\nu}F_{\lambda\sigma}F^{\lambda\sigma}) \,,  \\
T^\textrm{fluid}_{\mu\nu}&=& (\rho+p)u_\mu u_\nu+p\,g_{\mu\nu} \,, \\
J^\textrm{fluid}_\mu &=& \sigma u_\mu \,,
\end{eqnarray}
where $\sigma$ is the charge density of the fluid, $\rho$ is its energy density,
$p$ the pressure, and $u^\mu$ the four velocity, $u^\mu u_\mu=-1$.
The justification and limitations of the perfect fluid description are 
discussed in detail in \cite{Hartnoll:2010gu} and the same
considerations apply here. 

We look for static black brane solutions with planar 
symmetry,
\begin{equation}\label{ansatz}
ds^2=-f(v) dt^2 + g(v) dv^2 + \frac{1}{v^2}(dx^2+dy^2)\,,
\qquad
A = \frac{e}{\kappa} h(v) \, dt \,,
\end{equation}
where the radial coordinate goes from $v\rightarrow 0$ at the asymptotic
boundary to a constant value $v=v_0$ at the black brane horizon. We find it 
convenient to introduce a scale invariant variable $u=-\log(v/v_0)$, such that 
$u=0$ at the horizon and $u\rightarrow \infty$ at the boundary, and work with
rescaled fields,
\begin{equation}
\hat{f}=v_0^2 f,\quad \hat{g}=v_0^2 g, 
\quad \hat{h}=v_0 h, \quad \hat{p}=\kappa^2 p,
\quad \hat{\rho}=\kappa^2\rho ,\quad \hat{\sigma}=e\kappa\sigma \,.
\label{hatvars}
\end{equation}
The equations of motion (\ref{fieldeqs}) can then be expressed in a first
order form, convenient for numerical evaluation, 
\begin{eqnarray}
\frac{d\hat f}{du}+\frac{\hat{k}^2}{2}+\hat f(1-3e^{-2u}\hat g)&=&
e^{-2u}\hat f\hat{g}\hat p \,, \label{fequation} \\
\frac{d\hat k}{du}+\hat{k}&=&e^{-2u}
\left(\frac{1}{2}\hat{h}\hat{k}+\hat{f} \right)
\frac{\hat{g}\hat{\sigma}}{\sqrt{\hat f}} \,,\label{kequation} \\
\frac{1}{\hat f}\frac{d\hat f}{du}+\frac{1}{\hat g}\frac{d\hat g}{du}-4&=&
e^{-2u}\frac{\hat{g}\hat{h} \hat\sigma}{\sqrt{\hat f}} \,, \label{fgequation}
\end{eqnarray}
where $\hat k\equiv d\hat h/du$.
Following \cite{Hartnoll:2010gu}, we assume a free fermion
equation of state defined via 
\begin{equation}
\hat{\sigma}=\hat{\beta} \int_{\hat{m}}^{\hat\mu} 
d\varepsilon\,\varepsilon\sqrt{\varepsilon^2-\hat{m}^2} \,,\qquad
\hat{\rho}=\hat{\beta} \int_{\hat{m}}^{\hat\mu}   
d\varepsilon\,\varepsilon^2\sqrt{\varepsilon^2-\hat{m}^2}  \,,\qquad
-\hat p=\hat\rho -{\hat\mu} \hat\sigma \,,
\label{stateeq}
\end{equation}
where $\hat\beta$ is a coupling dependent dimensionless constant, 
$\hat m$ is proportional to the electron mass, 
$\hat{m}^2= \frac{\kappa^2}{e^2}m^2$, 
and the (rescaled) local chemical potential $\hat\mu$ is given by the 
background Maxwell gauge field in the tangent frame,
$\hat{\mu}\equiv\hat{h}/\sqrt{\hat f}$. 

As discussed in \cite{Hartnoll:2010gu}, there is a range of parameters,
\begin{equation}
e^2\sim \frac{\kappa}{L}\ll 1 \,,
\end{equation}
for which we can assume a classical bulk geometry with a non-trivial 
back-reaction due to the fermion fluid. If, at the same time, the Compton 
wavelength of the fermions is small compared to the AdS length scale,
\begin{equation}
mL\gg 1\,, 
\end{equation}
then we are also justified in taking spacetime to be locally
flat in the fermion equation of state. These conditions amount to the 
dimensionless parameters in the equation of 
state (\ref{stateeq}) taking order one values \cite{Hartnoll:2010gu},
\begin{equation}
\hat\beta\sim 1\,,  \qquad {\hat m}^2 \sim 1\,.
\end{equation}

The construction of the electron cloud geometry proceeds in a few
steps. First we solve the vacuum equations, with 
$\hat\sigma=\hat\rho=\hat p =0$, to find the charged AdS-RN black brane
solution inside the cloud,
\begin{equation}
\hat f=e^{2u}+\frac{{\hat q}^2}{2}e^{-2u}-(1+\frac{{\hat q}^2}{2})e^{-u} \,,\qquad
\hat g=\frac{e^{4u}}{\hat f}\,,\qquad 
\hat h=\hat q (1-e^{-u}) \,.
\label{rnsolution}
\end{equation}
The dimensionless constant $\hat q$ is proportional to the charge carried by 
the black brane and we have used the freedom to rescale the time coordinate $t$
to fix the overall normalization of $\hat f$. When the charge parameter is in the 
range $\hat q^2< 6$ the black brane is non-extremal with a non-degenerate horizon,
where the local chemical potential $\hat\mu$ vanishes. As we move away from the
horizon the chemical potential grows but remains too small to support a 
fermion fluid until
\begin{equation}
\hat\mu^2=\frac{\hat h^2(u)}{\hat f(u)} > \hat m^2\,.
\label{mucond}
\end{equation}
We will only consider non-vanishing fermion mass. For zero fermion 
mass the inner edge of the electron cloud reaches the horizon for all 
temperatures. The geometry, with the back-reaction from the fermion
fluid included, is then harder to determine and we will not consider this
case here.

\begin{figure} 
\begin{center}
\includegraphics[scale=1.1]{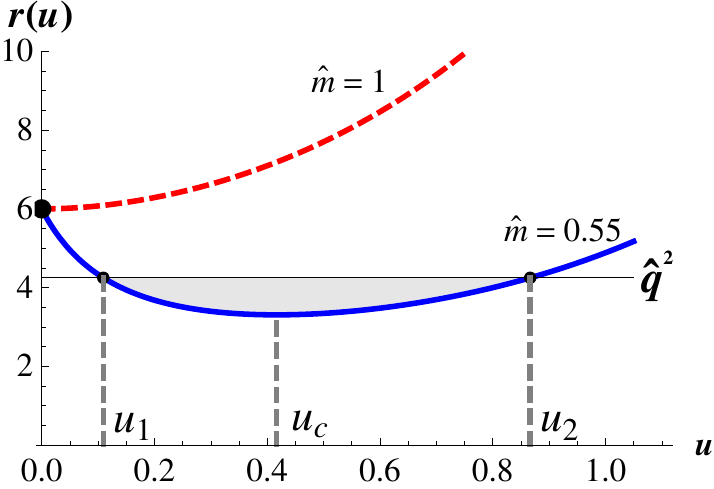} 
\caption{The auxiliary function $r(u)$ in (\ref{ucondition}), used in the
construction of the electron cloud solution, is plotted for $\hat m=1$ 
(dashed red) and $\hat m=0.55$ (solid blue).}
\label{fig:ucondition} 
\end{center}
\end{figure}

The condition (\ref{mucond}) is easily seen to be equivalent to 
\begin{equation}
\hat q^2>\frac{\hat m^2 e^u(e^{2u}+e^u+1)}{e^u-1+\frac{\hat m^2}{2}}
\equiv r(u) \,.
\label{ucondition}
\end{equation}
The right hand side is shown for two different values 
of $\hat m^2$ in Figure~\ref{fig:ucondition}. It can be read off from the figure that:
\begin{itemize}
\item
There cannot be any fermion fluid outside a non-extremal black brane if 
$\hat m^2\geq 1$, since in this case $r(u)> 6$ for all $u>0$. This restriction on 
$\hat m$ was already seen in \cite{Hartnoll:2010gu} as a condition for the 
existence of electron star solutions at zero temperature.
\item
For $\hat m^2< 1$ and a near-extremal black brane with $\hat q^2< 6$,
the condition (\ref{ucondition}) is satisfied within a finite interval $u_1<u<u_2$, 
indicated in the figure. The endpoints of the interval correspond to the inner and 
outer edges of the electron cloud in a ``probe'' approximation, where the 
back-reaction on the geometry due to the fermion fluid is ignored. 
\item 
For $\hat m^2< 1$ and $\hat q^2$ below a critical value (which depends on
$\hat m^2$), the condition (\ref{ucondition}) is not satisfied for any $u>0$. This 
implies there is a critical temperature above which there is only a black 
brane and no electron fluid in the bulk.
\end{itemize}

The second step in the construction of an electron cloud solution with back-reaction
included is to numerically integrate the field equations (\ref{fequation}) - 
(\ref{fgequation}) starting from the inner edge of the electron cloud. 
The functions $\hat f$, $\hat g$, $\hat h$, and $\hat k$ are continuous at the 
matching point $u=u_1$ and thus we can generate initial values for the numerical 
integration using the exact AdS-RN solution (\ref{rnsolution}) with $\hat q$ 
determined from (\ref{ucondition}) evaluated at $u=u_1$.

\begin{figure} 
\begin{center}
\includegraphics[scale=1]{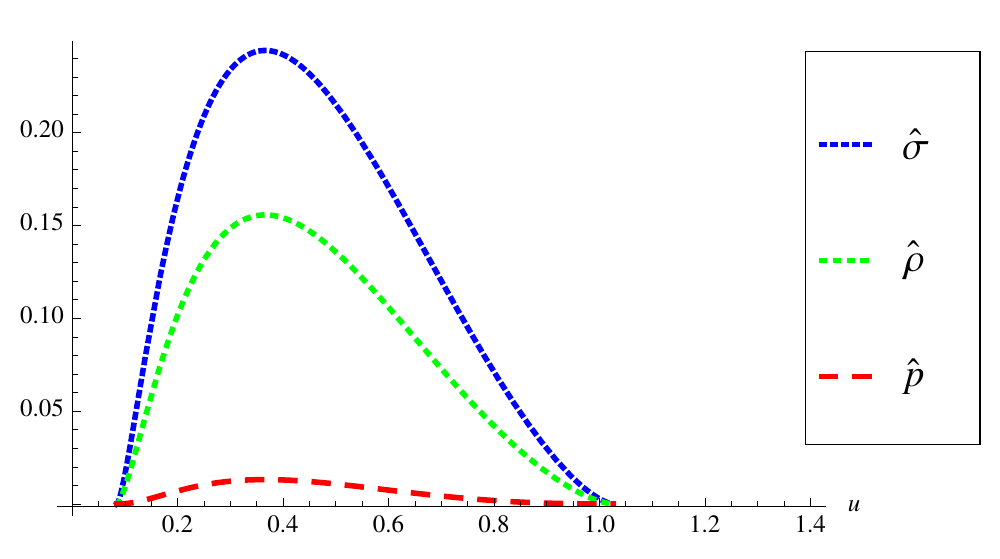} 
\caption{The radial profiles of the fluid variables $(\hat{\sigma},\hat{\rho},\hat{p})$ for
$\hat m=0.55$, $\hat\beta=10$, and $\hat q^2=4.49$.}
\label{fig:profiles} 
\end{center}
\end{figure}

The local chemical potential $\hat\mu$ goes to zero in the asymptotic 
$u\rightarrow\infty$ region and the numerical integration is terminated at a 
point $u=u_s$ where the condition (\ref{ucondition}) is no longer satisfied.
We find that the back-reaction of the fermion fluid on the geometry leads 
to $u_s>u_2$, and that this effect becomes more pronounced, $u_s\gg u_2$, 
at low temperature.
Figure~\ref{fig:profiles} shows numerical
results for the fluid variables $\hat\sigma$, $\hat\rho$, and $\hat p$ for
$\hat m=0.55$, $\hat\beta=10$, and $\hat q^2=4.49$. 

The third and final step in the construction is to obtain the spacetime geometry
outside the electron cloud by matching the numerical solution onto a charged 
black brane solution at $u=u_s$ in much the same
way as is done for electron stars in \cite{Hartnoll:2010gu}.
The exterior solution has the general form
\begin{equation}
\hat f=c_s^2e^{2u}+\frac{Q_s^2}{2}e^{-2u}-M_s e^{-u} \,,\qquad
\hat g=\frac{c_s^2 e^{4u}}{\hat f}\,,\qquad 
\hat h=\mu_s -Q_s e^{-u} \,,
\label{extsolution}
\end{equation}
where the subscript $s$ on the various constant parameters is a 
reminder that they are determined by matching onto a numerical solution
at $u=u_s$. We already fixed the overall scale of the time coordinate when
writing the inside black brane solution in (\ref{rnsolution}) so now there 
appears an extra parameter $c_s$ in $\hat f$. Also, since the external 
solution only extends to $u=u_s$ and not to an event horizon, we do not 
require the usual relationship between $M_s$, $Q_s$, and $\mu_s$ found 
for a vacuum AdS-RN black brane. The parameters in (\ref{extsolution}) 
are instead determined to be
\begin{eqnarray}
c_s^2&=&\hat f(u_s)\hat g(u_s) e^{-4u_s} \,,\label{eq:cs}\\
Q_s&=&\hat k(u_s)e^{u_s} \,,\label{eq:Qs}\\ 
\mu_s &=&\hat h(u_s)+\hat k(u_s) \,,\label{eq:mus}\\
M_s&=&\hat f(u_s)\hat g(u_s)e^{-u_s}
+\frac{1}{2}\hat k^2(u_s)e^{u_s}-\hat f(u_s)e^{u_s} \,.\label{eq:Ms}
\end{eqnarray}
These parameters refer to the rescaled fields in (\ref{hatvars}) while
the physical parameters appearing in an external AdS-RN solution 
with a canonically normalized time coordinate are given by
\begin{equation}
\mu= \frac{\mu_s}{ c_s v_0}  \,,\qquad
Q = \frac{Q_s}{c_s v_0^2}  \,,\qquad 
M = \frac{M_s}{c_s^2 v_0^3} \,. 
\label{physvars}
\end{equation}
Once the parameters of the external black brane solution have been 
determined for given values of $\hat m$, $\hat\beta$, and $\hat q$, 
standard methods can be used to obtain the free energy density as a 
function of temperature for these geometries. This will be carried out in 
Section \ref{freeenergy} below.

The next step is to determine the Hawking temperature of the electron cloud 
geometry, which is to be identified with the temperature in the boundary field theory.
The Hawking temperature is easily obtained from the behavior of the Euclidean 
metric near the horizon. One finds
\begin{equation}
\frac{T}{\mu}=\frac{6-\hat{q}^2}{8\pi\mu_s}\,,
\label{hawkingtemp}
\end{equation}
where we have again divided by the physical chemical potential $\mu$ in 
order to have a dimensionless quantity to work with.

\begin{figure} 
\begin{center}
\includegraphics[scale=1.4]{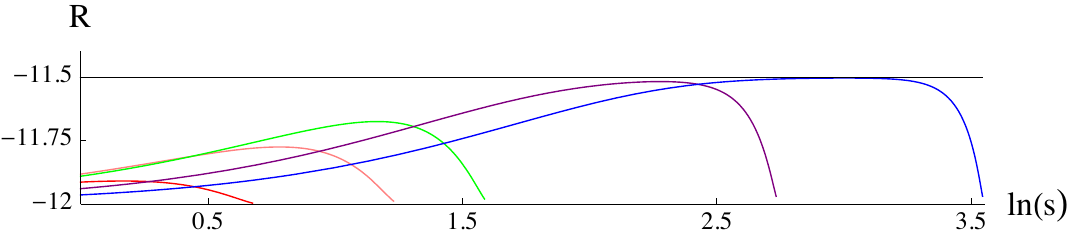} 
\caption{The curvature scalar $R$ versus the proper distance $s$ measured from the outer edge of the electron cloud for $\hat m=.55\,, \hat \beta=10$ and for $T/\mu$ values $.55, .22,
9.4 \times 10^{-2},  9 \times10^{-4}$ and  $1 \times 10^{-5}$. The curves extend further to the right with decreasing temperature. The value for $R$ in the Lifshitz region deep inside an electron star with the same $\hat m$ and $\hat \beta$ is shown as a horizontal line for reference. }
\label{fig:curvature} 
\end{center}
\end{figure}

In the limit of zero temperature we expect to recover the electron star 
geometry of \cite{Hartnoll:2010gu}. In this case the radial coordinate 
extends to $u\rightarrow -\infty$ and the metric exhibits Lifshitz scaling 
in the deep interior.

In Figure~\ref{fig:curvature} one can see that the curvature scalar $R$ as a function of the proper distance $s$ measured from the outside of the electron cloud is approaching the expected asymptotic value for a Lifshitz geometry as we lower the temperature $T$ before plunging to $R=-12$ at the horizon. From $R$ evaluated on the solution and the metric, one confirms the correct dynamical exponent, e.g. $z=5.75466$ when $\hat m=.55\,$ and $\hat \beta=10$.

\begin{figure} 
\begin{center}
\includegraphics[scale=1.2]{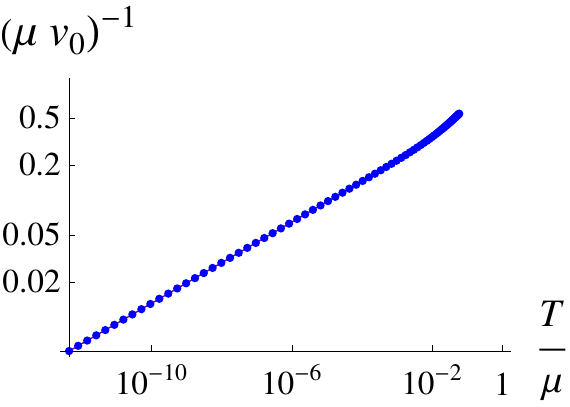} 
\includegraphics[scale=1.2]{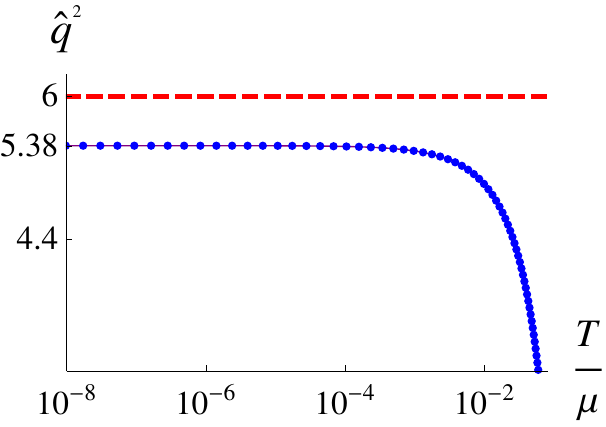}
\caption{On the left we plot the horizon radius $v_0^{-1}$ as a function of the temperature, $T$. On the right we show the charge parameter of the inside black brane solution \eqref{rnsolution}, $\hat q^2$, versus $T$ for $\hat m=.55$ and $\hat \beta=10$. For reference we plot the extremal value $\hat q^2=6$.}
\label{fig:qhatT} 
\end{center}
\end{figure}

On the left in Figure~\ref{fig:qhatT}, we see that the horizon radius vanishes as the temperature is lowered indicating that the horizon recedes from the AdS boundary. Furthermore by fitting the low temperature data in the figure, we find that 
\begin{equation}
{T\over\mu} \propto \left({1\over \mu v_0}\right)^z\,,
\end{equation}
where $z$ matches the appropriate expected Lifshitz exponent to a very high precision. This is further evidence that the anisotropic scaling found inside the electron star is recovered by our electron cloud solutions at low temperature.

The plot on the right in Figure~\ref{fig:qhatT} shows that the zero temperature limit in the boundary theory is in fact not obtained by approaching an extremal interior black brane, which would have $\hat q^2 =6$. This is at first sight counterintuitive but can ultimately be traced to the focusing effect that the electron fluid has on the geometry outside the black brane. If one were to try to construct an electron cloud solution starting with a value of $\hat q^2$ that is closer to $6$, than what is seen in the figure the metric would collapse to a curvature singularity at a finite proper distance outside the horizon and the solution would never reach an asymptotic AdS region.


\section{Free energy}
\label{freeenergy}

Further evidence that the electron cloud
solution is the proper finite temperature extension of an electron star comes from 
comparing free energy densities.
We obtain the free energy by evaluating the on-shell 
Euclidean action of the bulk system, including the usual Gibbons-Hawking 
boundary term \cite{Gibbons:1976ue} and boundary counterterms required
for regularization \cite{Henningson:1998gx,Balasubramanian:1999re}. 
A bulk action for the charged electron fluid also needs to be included, as
described in \cite{Hartnoll:2010gu}. The full bulk 
action turns out to be the integral of a total derivative, and when combined with 
the appropriate boundary terms, it gives a simple result for the free energy 
density,
\begin{equation}\label{eq:consentropy}
F = M-\mu\, Q \, -sT \,,
\end{equation}
where $s$ is the Bekenstein-Hawking entropy density.\footnote{In an earlier version of the paper the $sT$ term was missing from the expression for the free energy density.  This led to an incorrect characterization of the phase transition.} This can be simplified by using the thermodynamic relation 
\begin{equation}
\label{eq:consentropy2}
\frac{3}{2} M-\mu\, Q-s\, T=0\,,\end{equation} giving
\begin{equation}
F = -\frac{M}{2}\,.
\end{equation}  The relation \eqref{eq:consentropy} follows from the radial conservation of the quantity 
\begin{equation}
D = \frac{e^{3u}}{\sqrt{\hat{f}\hat{g}}}\left(-2\hat{h}\hat{k}-2\hat{f}+\frac{d \hat{f}}{d u}\right)\,.\end{equation}
By using the equations of motion \eqref{fequation} - \eqref {stateeq}, it is straightforward to check that $\frac{d D}{d u} = 0$ and one then evaluates $D$ at the horizon and at the $u\rightarrow \infty$ boundary to obtain \eqref{eq:consentropy2}.  

Using (\ref{physvars}), the free energy density can be re-expressed in terms  of output parameters from our numerical evaluation,
\begin{equation}
F= -\frac{1}{2}\frac{ M_{s} }{c_s^2 v_0^3} \,.
\end{equation}
The factor of $v_0^3$ in the denominator tells us that we should instead work
with the dimensionless quantity 
\begin{equation}
\frac{F}{\mu^3}= -\frac{1}{2} \frac{c_s M_{s}}{\mu_s^3}\,,
\label{dimensionlessF}
\end{equation}
when comparing free energy densities.

\begin{figure} 
\begin{center}
\includegraphics[scale=1]{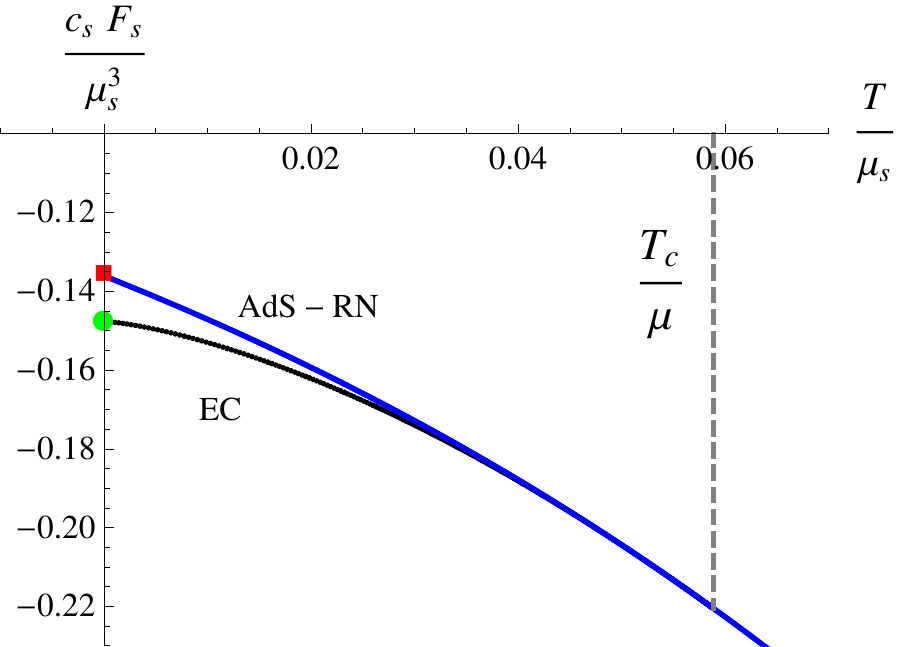} 
\caption{The free energy densities of AdS-RN black brane and electron cloud 
solutions for $\hat{\beta}=10$ and $\hat{m}=.55$. In addition the free energy for an extremal black hole and the electron star 
solution of \cite{Hartnoll:2010gu} are shown with a red box and a green  dot, 
respectively.}
\label{fig:fenergy} 
\end{center}
\end{figure}

In Figure \ref{fig:fenergy} we use these dimensionless variables to compare 
the free energy densities of various geometries for a typical case when 
$\hat{m}=.55$ and $\hat{\beta}=10$, holding the chemical potential fixed.  
One readily sees that the electron cloud solution is preferred over the black 
brane solution up to the point where the local chemical potential is too low 
to support any fluid.  Beyond this point the only solution is an AdS-RN black 
brane.  At low temperatures, on the other hand, the free energy density of 
the electron cloud geometries approaches that of the corresponding electron 
star.  

In addition to the low-temperature regime, it is also interesting to ask about the 
nature of the transition to the AdS-RN black brane solution at higher temperatures.  
To address this issue, we consider the difference in energy densities between an electron cloud 
solution just below the critical temperature $T_c$ and an AdS-RN black brane
solution,
\begin{equation}
\Delta\left(\frac{F}{\mu^3}\right) 
\equiv \left(\frac{ F}{\mu^3}\right)_{AdS-RN} -\left(\frac{ F}{\mu^3}\right)_{EC}\,,
\end{equation}  at the same value of $T/\mu$.
\begin{figure} 
\begin{center}
\includegraphics[scale=1.1]{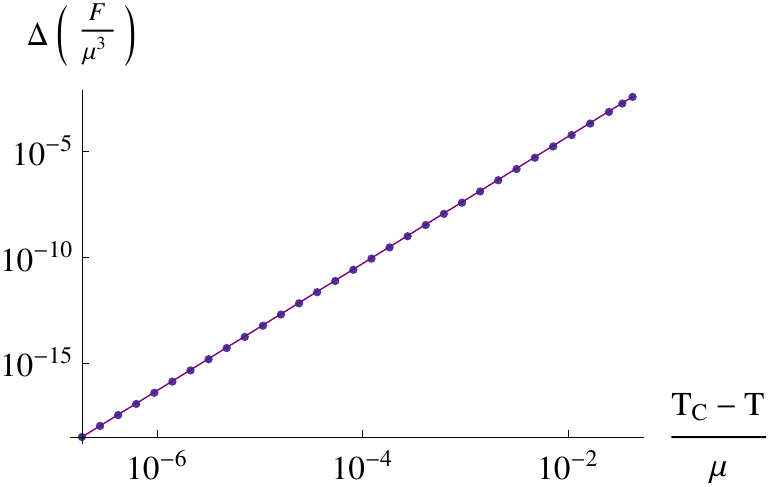} 
\caption{The difference between the black brane and electron cloud free 
energies near the phase transition temperature at $\hat{\beta}=10$ and $\hat{m}=.55$.}
\label{fig:energydiff} 
\end{center}
\end{figure}
Figure \ref{fig:energydiff} shows a log-log plot of this difference near the critical point where one loses the cloud solution at $T_{c}/\mu=0.058868$ for  
$\hat{m}=.55$ and $\hat{\beta}=10$.  The solid curve in Figure \ref{fig:energydiff} is a straight line of slope $3$ giving numerical evidence of a third order phase transition where 
\begin{equation}
\Delta\left(\frac{F}{\mu^3}\right)=\mathcal{O}\left(\frac{T_c-T}{\mu}\right)^3\,.\end{equation}  This feature was observed numerically in \cite{Hartnoll:2010ik}, and those authors also gave a simple analytic argument for this behavior.

\section{Electric conductivity}

The finite temperature AC conductivity at zero momentum can be computed 
in an analogous way as was done for the zero temperature electron star
in \cite{Hartnoll:2010gu}. In fact, the finite temperature computation is more 
standard since in this case the ingoing boundary conditions \cite{Son:2002sd} 
for the fluctuations in the gauge field are imposed at a smooth black brane horizon
rather than at the (mildly) singular Lifshitz horizon inside an electron star.

The background is perturbed, assuming a time dependence of the form 
$e^{-i \omega t}$, and the resulting equations are linearized.
To get a closed system of equations, the following perturbations are needed
\begin{equation}
\hat A_x = \delta \hat A_x(u) e^{-i \omega t} \,,\qquad
\hat g_{tx} = \delta \hat g_{tx}(u) e^{-i \omega t} \,,\qquad 
\hat u_x = \delta \hat u_x(u) e^{-i \omega t} . 
\end{equation}
This leads to a system of four first order differential equations
\begin{eqnarray}
\delta \hat A_x + \frac{\hat h}{\hat f}\, \delta \hat g_{tx}+e^{2 u}{\hat \mu}\,\delta \hat u_{x}			
&=&	0,	\label{u_x}	\\
{d\delta \hat g_{tx}\over du}-2 \delta \hat g_{tx}+2 \,{d \hat h\over d u}\, \delta \hat A_x 					
&=&	0,	\label{g_tx}	\\
e^{u}{d \delta \hat A_x\over du}+ \sqrt{\frac{\hat g}{\hat f}}\delta \hat B_{x} 				
&=&	0,	\label{A_x}	\\
e^{u}\,{d \delta \hat B_{x}\over du} + \left[{1\over \sqrt{\hat f \hat g}} \left(2 e^{2 u} {d \hat h\over d u}^2-\omega_s^2 \hat g\right)
+ \frac{\hat f\hat \sigma \sqrt{\hat g}}{\hat h} \right] \delta \hat A_x 	
&=&	0,	\label{B_x}
\end{eqnarray}
where $\omega_s$ is defined in terms of the canonical normalized frequency $\omega$ as $\omega_s=\, c_s \,v_0\, \omega$. Equation (\ref{A_x}) can be regarded as the definition of the auxiliary function $\delta \hat B_x$.
We note that (\ref{A_x}) and (\ref{B_x}) form a closed system involving only $\delta \hat A_x $ and $\delta \hat B_x$.

At the horizon, ingoing boundary conditions are imposed \cite{Son:2002sd}.
This implies $\delta \hat A_x \to u^{-\frac{i \omega}{4 \pi T}}$ and $\delta \hat B_x \to i\omega u^{-\frac{i \omega}{4 \pi T}}$
as $u \to 0$ and $T$ being the Hawking temperature of the AdS-RN black brane solution (\ref{rnsolution}).
At the AdS boundary, where the background is of the form (\ref{extsolution}), the behavior of those functions is
\begin{eqnarray}
\delta \hat A_x &=&	 \hat A_x^{(0)} + \hat A_x^{(1)} e^{-u} + \cdots \,,	\label{A_x_bdry}	\\
\delta \hat B_x &=& \hat B_x^{(0)} + \hat B_x^{(1)} e^{-u} + \cdots \,.	\label{B_x_bdry}
\end{eqnarray}
The coefficients $\hat A_x^{(i)}$ and $\hat B_x^{(i)}$ are connected, \emph{e.g.} $\hat B_x^{(0)} = c_s \hat A_x^{(1)}$.
This relation can be used to express the conductivity as 
\begin{equation}
\sigma = -\frac{i}{\omega_s}\frac{ \hat B_x^{(0)}}{ \hat A_x^{(0)}} \,, \label{conductivity}
\end{equation}
which is manifestly invariant under the rescaling described in the previous sections.

\begin{figure} 
\begin{center}
\includegraphics[scale=.8]{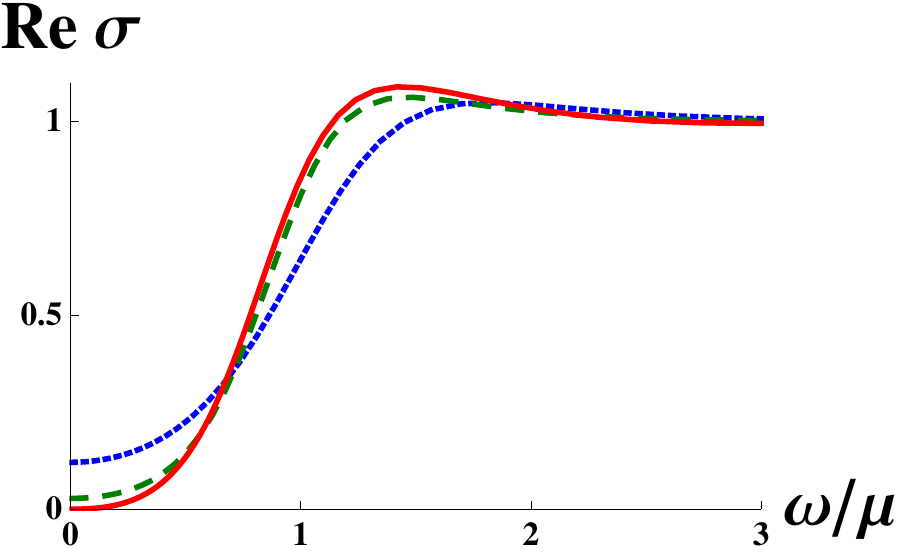} 
\includegraphics[scale=.8]{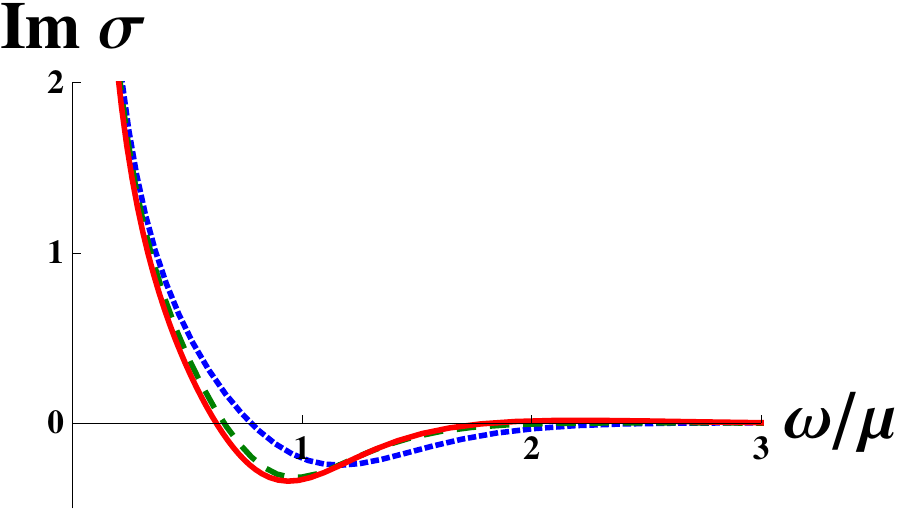} 
\caption{Real and Imaginary part of the conductivity for $\hat m=0.55, \hat\beta=10$. The curves blue dots, dashed green and solid red correspond to temperature value $T/T_C = 2, 1$ and $0.162$, respectively. $T_C/\mu = 0.05887$ denotes the critical temperature where the phase transition occurs. Curves for lower values of $T$ are almost indistinguishable from the solid red curve.}
\label{fig:conductivity} 
\end{center}
\end{figure}

Our results for the conductivity are obtained by numerics.
This is achieved by integrating out from the horizon in the background of an AdS-RN solution, as already indicated, until the inner edge of the electron shell is reached.
There, $\delta \hat A_x$ and $\delta \hat B_x$ need to be continued smoothly into a solution of (\ref{A_x}) and (\ref{B_x}) with the electron cloud solution as background.
At the outer edge, a second matching to the exterior solution must occur.
Finally, the coefficients $\hat A_x^{(0)}$ and $\hat B_x^{(0)}$ can be read off at the boundary and plugged in to (\ref{conductivity}).

\begin{figure} 
\begin{center}
\includegraphics[scale=.7]{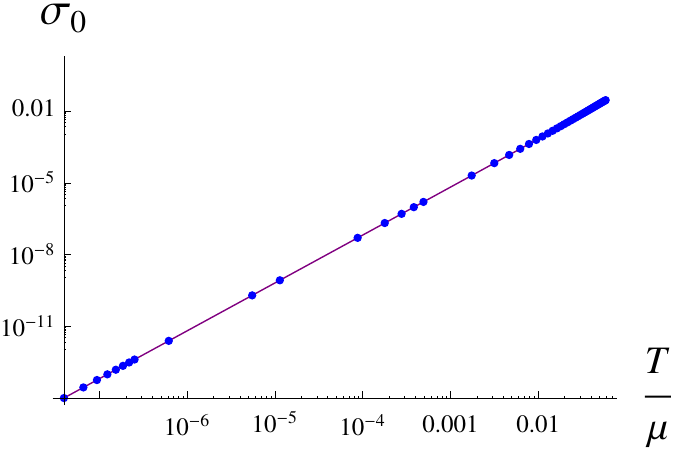}
\includegraphics[scale=.7]{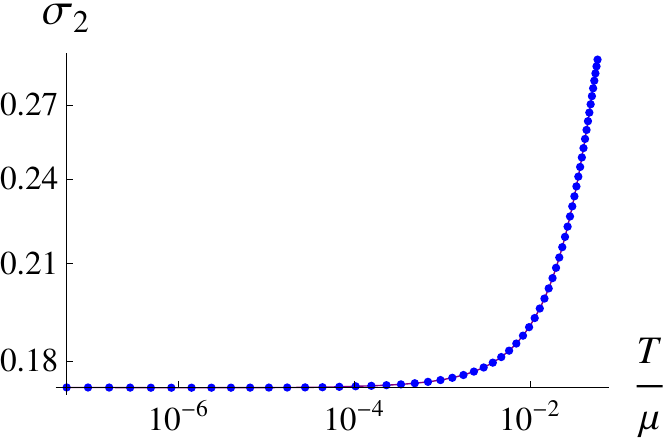}
\includegraphics[scale=.7]{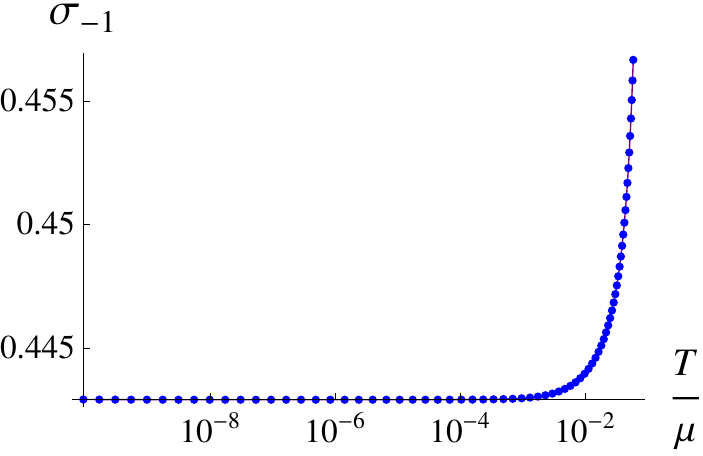}
\caption{Numerical results for the coefficients $\sigma_0,\sigma_2$ and $\sigma_{-1}$ in the electrical conductivity in equation \eqref{sigma_plot}.
}
\label{fig:sigmas} 
\end{center}
\end{figure}

A plot of the conductivity can be seen in Figure \ref{fig:conductivity}.
The pole in the imaginary part, as usual, indicates the presence of a delta peak in the real part.
The offset in the conductivity goes rather quickly to zero once the electron cloud is in place.
This is also shown in Figure \ref{fig:sigmas}. Parameterizing the real and imaginary part of the conductivity as 
\begin{equation}
\label{sigma_plot}
\mbox{Re}\,\sigma \sim \sigma_0 + \sigma_2 \left(\frac{\omega}{\mu}\right)^2 \,,
\qquad
\mbox{Im}\,\sigma \sim \sigma_{-1}\frac{\mu}{\omega}
\end{equation}
 for small $\omega$, it can be seen from our numerical data that $\sigma_2$ and $\sigma_{-1}$ level off for small temperatures and $\sigma_0$ decreases like $T^2$ in this limit. This is consistent  with the zero-temperature result in \cite{Hartnoll:2010gu}, where $\sigma_0$ is not present, and with $\omega/T$ scaling at low temperatures. Above the critical temperature the background is an AdS-RN black brane and the calculation of the conductivity reduces to the one described in~\cite{Hartnoll:2009sz}. 

\section{Discussion}

One of the most interesting recent developments in holographic model building is the 
observation that at low temperatures it can be energetically favorable for black holes 
in AdS space to eject their charge in the form of matter "hair" 
\cite{Hartnoll:2009ns, Hartnoll:2008kx, Gubser:2008pf, Horowitz:2009ij, Gauntlett:2009dn}.  
In the boundary theory this hair may give rise to many interesting features 
including spontaneous symmetry breaking \cite{Gubser:2008px},  dynamical 
critical exponents \cite{Hartnoll:2010gu} and non-fermi liquids 
\cite{Hartnoll:2009ns,Hartnoll:2010xj}. In general, this may allow one to 
use holographic techniques to study condensed matter systems not amenable 
to other theoretical tools.  

In this paper we explored the finite boundary temperature generalization of the electron star configuration described in \cite{Hartnoll:2010gu}.  The electron star solution is a zero temperature model for a quantum phase transition displaying dynamical critical exponents as well as non-fermi liquid features.  In the bulk the configuration is that of a zero temperature ideal charged fermion fluid.  In the deep interior this fluid has an asymptotic Lifshitz scaling symmetry.  As one approaches the boundary, the gauge potential is screened by the charged fluid.  Eventually, the local chemical potential falls below the fermion mass causing the fluid density to vanish.  In the asymptotic region one is left with an AdS-RN geometry.

At finite boundary temperatures there is instead a cloud-like configuration, with the electron fluid hovering outside an AdS-RN black brane geometry.  This configuration is only possible when the fluid and black brane have same sign charges such that electrostatic repulsion balances gravitational attraction. Within the fluid and its exterior, the electron cloud solution is similar to the electron star.  The gauge field is screened and eventually the fluid can no longer be supported.  We found that as one lowers the temperature, for fixed chemical potential one smoothly obtains the free energy of the electron star solution at zero temperature.  

In the other extreme, beyond a critical temperature the local chemical potential is always lower than the fermion mass and no fermion fluid is supported.  In this case we are left with an AdS-RN black brane geometry.  We studied the transition numerically and found that it is a third order phase transition, as pointed out in \cite{Hartnoll:2010ik}.  

We can summarize the phase diagram as follows, in the high temperature regime there is an AdS-RN black brane.  As one lowers the temperature it is favorable for the black brane to expel some of its charge in the form of an electron cloud hovering over the horizon.  As one cools the system further, the interior black brane carries less and less charge and has shrinking area.  Finally, at zero temperature the black brane is gone and the fluid takes the form of an electron star.

In addition to studying electron cloud thermodynamics we also computed the conductivity, finding that it is nicely consistent with the electron star results of \cite{Hartnoll:2010gu} at low temperature and with the AdS-RN black brane at high temperature. 

In \cite{Hartnoll:2010xj} it was argued that if the fluid in an electron star experiences a local temperature and magnetic field, it is possible to detect a Fermi surface evidenced by Kosevich-Lifshitz oscillations. It should also be possible to see such oscillations in electron clouds.

A major challenge for the construction we are working with is how to
interpret the electron fluid directly in terms of operators in the boundary 
field theory.  It would be interesting to see how the work of \cite{Arsiwalla:2010bt} 
may be extended for charged fluids in order to further characterize the nature 
of the underlying quantum critical point.

\acknowledgments

We thank S.~A.~Hartnoll and K.~Schalm for helpful discussions. This work was supported in part by the Icelandic Research Fund and by the 
University of Iceland Research Fund.

\end{document}